\documentclass{cup-hpl}
\usepackage[none]{hyphenat}
\usepackage{orcidlink}
\usepackage{siunitx}
\usepackage{booktabs}
\usepackage{graphicx} 
\usepackage{subcaption}
\usepackage[nice]{nicefrac}
\usepackage{amsmath}
\usepackage{microtype}
\usepackage{xcolor}
\usepackage{float}

\begin{document}

\newtheorem{theorem}{Theorem}

\shorttitle{Analysis of Seismic Influences at EuXFEL}
\shortauthor{Arne Grünhagen et al.}

\title{Impact of Seismic Activities on the Optical Synchronization System of the European X-ray Free-Electron Laser}

\author[1,2]{Arne Grünhagen \orcidlink{0009-0001-5541-4011} \corresp{{arne.gruenhagen@desy.de}}}
\author[1]{Maximilian Schütte}
\author[1]{Thorsten Lamb}
\author[1]{Sebastian Schulz \orcidlink{0000-0002-6306-703X}}
\author[1,3]{Annika Eichler}
\author[2]{Marina Tropmann-Frick}
\author[3]{Görschwin Fey}
\author[1]{Holger Schlarb}

\address[1]{Deutsches Elektronen-Synchrotron DESY, Germany}
\address[2]{Hamburg University of Applied Sciences, HAW, Germany}
\address[3]{Hamburg University of Technology, TUHH, Germany}

\begin{abstract}
This study investigates the influence of seismic activities on the optical synchronization system of the European X-ray Free-Electron Laser. We analyze the controller I/O data of phase-locked-loops in length-stabilized links, focusing on the response to earthquakes, ocean-generated microseism and civilization noise. By comparing the controller data with external data, we were able to identify disturbances and their effects on the control signals. Our results show that seismic events influence the stability of the phase-locked loops. Even earthquakes that are approximately \qty{5000}{\km} away cause remarkable fluctuations in the in-loop control signals. Ocean-generated microseism in particular has an enormous influence on the in-loop control signals due to its constant presence. The optical synchronization system is so highly sensitive that it can even identify vibrations caused by civilization, such as road traffic or major events like concerts or sport events. The phase-locked loops manages to eliminate more than \qty{99}{\percent} of the existing interference.  
\end{abstract}


\maketitle

\textcolor{black}{
\section{Introduction}}
The European X-ray Free-Electron Laser (EuXFEL) is a state-of-the-art research facility in Hamburg capable of generating X-ray flashes as short as a few femtoseconds\cite{sobolev2020megahertz}. These ultra-fast flashes enable groundbreaking studies of atomic structures and dynamics on extremely short time scales. The EuXFEL, spanning \qty{3.4}{\km} in a tunnel located \qtyrange{6}{30}{\m} underground, includes an injector, accelerating modules, diagnostics, undulators, and an underground experimental hall for pump-probe experiments. Central to the precision and functionality of the facility is the optical synchronization system, which ensures temporal coherence across all components.

The optical synchronization system\cite{schulz2015femtosecond} is capable of achieving femtosecond-level precision by distributing a phase-stable optical reference signal via a network of lasers and optical fibers. This reference signal, generated by an optical Main Laser Oscillator (MLO) phase-locked to a \qty{1.3}{\giga\hertz} main RF oscillator\cite{zembala2014master}, is distributed to \num{20} links in a free-space distribution network and delivered to various subsystems through optical fibers. Maintaining stable optical path lengths between the MLO and synchronized clients is essential for preserving phase coherence, a task achieved by Link Stabilization Units (LSUs) that compensate for fiber length variations in a Phase-Locked Loop (PLL).
\textcolor{black}{
Seismic activity, whether from natural events like earthquakes, ocean-generated microseism, or human activities, poses a significant challenge to the optical synchronization system. Vibrations or tunnel deformations caused by seismic waves alter the optical path length, inducing phase shifts in the synchronization signal. These effects can disrupt the stability of the phase-locked signals, compromising the precision of the system. Therefore, understanding how seismic disturbances affect the synchronization system and evaluating the suppression mechanisms are critical for ensuring its reliability.
}
This work systematically analyzes the influence of ground motion on the fiber-based synchronization system of the EuXFEL. We focus on the effects of global earthquakes, ocean-generated microseism, and civilization activities on the in-loop signals of the optical synchronization system. \textcolor{black}{To characterize these influences, we employ time-domain filtering techniques to segment sequential discrete time-based measurements and perform subsequent noise analysis. Additionally, we use Power Spectral Density (PSD) estimations with the Welch method\cite{1161901} for frequency analysis, which helps identify and quantify frequency components of the seismic disturbances. By applying these methods, we evaluate the system's response to seismic disturbances and assess its effectiveness in suppressing noise. This approach provides a detailed understanding of the system’s resilience and highlights the broader implications for synchronization systems in similar facilities.
}
This work is structured as follows: In Section~\ref{sec:rel-work} we refer to other work that also deals with external disturbances on large scale facilities. In Section~\ref{sec:technicals} we give an overview about the optical synchronization system of the EuXFEL and Section~\ref{sec:method} provides a detailed description of how we use this system to analyze and categorize seismic effects. Section~\ref{sec:results} describes the results of the analysis and in Section~\ref{sec:conclusion} the work is summarized and an outlook on future work is given.

\section{Related Work}
\label{sec:rel-work}

The effects of seismic activities on high-precision scientific instruments have been intensively investigated in various research areas. In the context of large-scale particle accelerators such as the EuXFEL, studies dealing with the stability and precision of experiments under seismic influences are particularly relevant. 

Due to the high stability and performance requirements of the Large Hadron Collider (LHC) for the beam orbit, there are several studies analyzing the effects of ground motion on the accelerator. In \cite{vos1995ground} and \cite{steinhagen2005analysis} the impact of seismic activities on the predecessors of the LHC, the Large Electron-Positron (LEP) collider and the Super Proton Synchrotron (SPS), was analyzed. To this end, seismic measurements from geophones were compared with beam motions from SPS and LEP. They show that seismic disturbances with a frequency above \qty{1}{\hertz} have only a negligible influence on the orbit. The beam position is mainly influenced by random ground motion, which can be seen below \qty{1}{\hertz}. In \cite{collette2010seismic}, the authors present recent ground motion measurements in the LHC tunnel. Using these measurements, they describe how they updated models of vertical and lateral ground motions. A dynamic model for linear accelerators was developed that can include local excitation sources and evaluate the seismic response of the linear accelerator.

The authors of \cite{scislo2022high} investigate the impact of earthquake swarm events on the LHC tunnel. Alarm thresholds for vertical ground motion were established based on proton loss and loss of luminosity. It is shown how often individual earthquakes as well as earthquake swarms exceed these alarm thresholds. This warning system is used because excessive ground motion can lead to serious safety problems due to the high beam and collision energies and the energy stored in the magnets.

Prior to the upgrade from the LHC to the High Luminosity Large Hadron Collider (HL-LHC), seismic sensors were installed near the Atlas and CMS experiments to monitor ground motions near the experiments during normal operation as well as during the construction of the upgrade from the LHC to the HL-LHC. In \cite{SCHAUMANN2023168495,gamba2018estimated} the influences are summarized. In particular, the influence of ground motions on orbital perturbations, beam losses and losses of brightness are analyzed. The analyses paired with simulations show that the HL-LHC is about twice as sensitive to ground motion as the LHC. This is already of such a magnitude that the resulting beam losses could lead to frequent beam shutdowns.

In \cite{simos2019nsls} the authors analyze the ground movements and the associated challenges for the electron beam of the National Synchrotron Light Source II (NSLS-II) at Brookhaven National Laboratory. In particular, the influence of different soil conditions on the attenuation of ground movements is analyzed. As sources of disturbance, a distinction is made between natural phenomena (e.g.,\,microseism and earthquakes) and civilization-made disturbances (e.g.,\,car traffic, train traffic, industrial machinery). Based on these analyses, proposals were made for the final design of the NSLS-II so that seismic disturbances are passively attenuated in the best possible way.

The authors of \cite{scislo2022retraction} show the influence of the reduction of human activity during the COVID-19 lockdown period on ground vibrations near the LHC. While human activity decreased, the strength of ground vibrations also decreased. However, the difference in vibration strength caused by human activity had no measurable influence on the stability of the beam.

The publications mentioned so far use seismometers and geophones to determine the ground movements and thus the influence on the particle accelerators. In addition to these classic methods, distributed acoustic sensing (DAS) is a promising technology that can be used to measure vibrations and thus ground movements along an optical fiber. This technology was used for several publications \cite{williams2019distributed,lin2024monitoring,sladen2019distributed} to identify similar ground movements as the conventional measurement methods with seismometers, but along the entire optical fiber and not just at individual measurement points. This measurement method is also currently being installed at DESY \cite{wavehamburg2024}.
\textcolor{black}{
The authors of \cite{noe2023long} present a method for earthquake detection using fiber optic cables based on active phase noise cancellation (PNC). This approach uses existing phase-stabilized networks without additional hardware and is compatible with cables longer than \qty{1000}{\kilo\meter}. Using data from a magnitude \num{3.9} earthquake and a \qty{123}{\kilo\meter} fiber optic link, they show that PNC sensor technology provides accurate seismic data and is therefore suitable for earthquake monitoring.
}

Unlike the systems described, whose primary task is to measure ground motion, we analyze the EuXFEL optical synchronization system, whose primary task is to synchronize components of the EuXFEL over a length of up to \qty{3.5}{\km}. With our methodology, we can precisely identify how different types of seismic disturbances affect the system's performance, providing a detailed understanding of the relationship between seismic activity and synchronization stability.

\section{Technical Background}
\label{sec:technicals}
\subsection{Optical Synchronization System of the EuXFEL}
The optical synchronization system (see Figure~\ref{fig:lbsyncscheme}) of the EuXFEL is used to synchronize various instruments and devices within the EuXFEL on a femtosecond scale. The main components of this system are:

\begin{figure*}[b]
  \includegraphics[width=\textwidth]{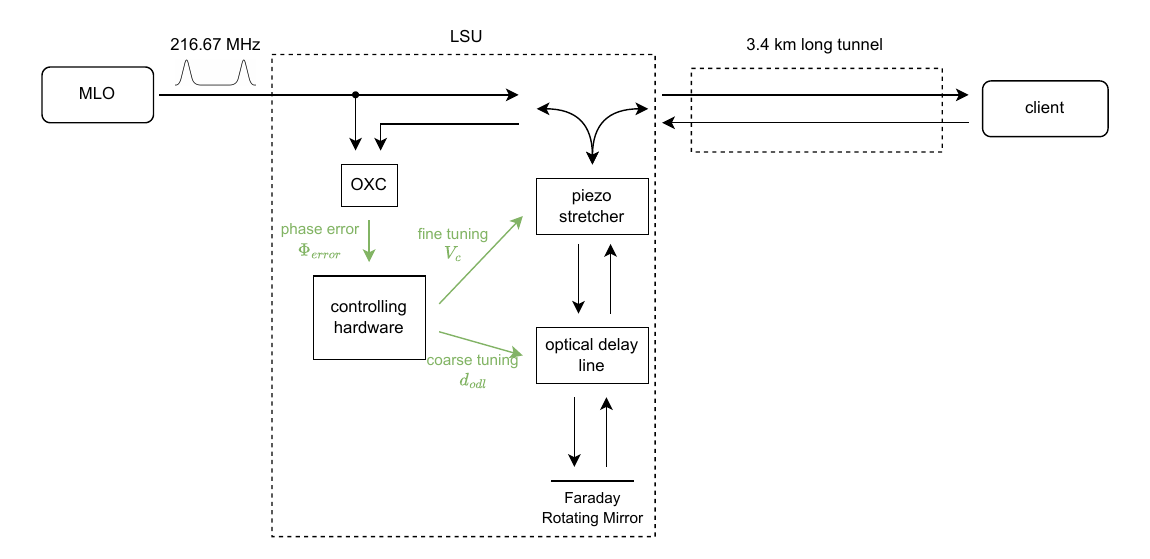}
  \caption{Schematic overview of a single link stabilizing unit of the optical synchronization system of the EuXFEL.}
  \label{fig:lbsyncscheme}
\end{figure*}

\begin{description}
    \item[Main Laser Oscillator (MLO)]
    The optical synchronization system contains two redundant main laser oscillators (MLO) which are phase-locked to the \qty{1.3}{\giga\hertz} RF signal of the main RF oscillator, both emitting a laser pulse train with a pulse repetition rate of \qty{216.67}{\mega\hertz} and a pulse duration of \qty{200}{\femto\second}. Phase locking is achieved by a Proportional-Integral (PI) control loop. This control loop ensures that the phase of the MLO is precisely matched to the reference phase of the main RF oscillator.

    \item[Synchronized clients]
    Various clients such as electron bunch arrival time monitors\cite{steffen2019electro}, RF re-synchronization modules\cite{lamb2011femtosecond}, and external laser systems are phase-locked to the optical synchronization system. In addition, there is a Secondary Laser Oscillator (SLO) in the experimental hall, which replicates the signal from the MLO and is used for the synchronization of components at the end of the EuXFEL, such as the pump-probe lasers.

    \item[Links]
    The connection between a synchronized end-station and the optical reference laser (MLO/SLO) is referred to as a link. The optical pulse train from the MLO is distributed into several links, with each link transporting the signal to a synchronized component via optical fibers. It is essential that the optical length of these links is actively stabilized. This stabilization is carried out by the Link Stabilization Unit (LSU). The optical pulse train from the MLO is transported via the optical fibers to the synchronized clients, where a partially reflecting mirror reflects part of the signal back to the LSU. Within the LSU the reflected pulse is overlapped in an optical cross-correlator (OXC) with a reference pulse. The temporal overlap is stabilized by using a PI controller to actively compensate the link length variations. Motorized optical delay lines are used for coarse tuning and piezoelectric fiber stretchers for fine tuning. These link stabilization units are intended to compensate for all occurring changes in length on the link fiber. These length changes are caused, for example, by changes in temperature and humidity in the tunnel. Seismic activity around the accelerator tunnel affects the position of the physical endpoints of the accelerator\cite{genthe2024impact}, leading to variations in the length of link fibers. This effect is measured and compensated by the LSU.
\end{description}

\subsection{Seismic Activities}
\label{sec:seismics}
A comprehensive understanding of seismic activity is crucial in order to understand what can be expected at the EuXFEL site. We distinguish between earthquakes, ocean-generated microseism and civilization-made ground motions. The principles of seismic activities described in the following are based on \cite{shearer2019introduction,longuet1950theory}.

\subsubsection{Earthquakes} are sudden discharges of energy that are usually caused by tectonic plate movements. This energy propagates in the form of seismic waves. The main types of seismic waves include body waves and surface waves. Body waves propagate through the earth's interior and are categorized into two main types.

\begin{description}
    \item [P-waves] (primary waves) are compression waves that propagate through alternating phases of compression and expansion. They have a high speed and a low amplitude. P-waves are usually in the frequency range from \qty{0.1}{\hertz} to \qty{100}{\hertz}.
    \item [S-waves] (secondary waves) are transverse waves in which particles move perpendicular to the direction of wave propagation due to shear displacements. They move more slowly than P-waves and often have a higher amplitude, which leads to stronger ground shaking. S-waves are typically in the frequency range from \qty{0.1}{\hertz} to \qty{50}{\hertz}.
\end{description}

Surface waves propagate along the earth's surface and occur when body waves reach the surface. They are divided into two main types:

\begin{description}
    \item[Rayleigh waves] are surface waves that cause an elliptical rolling motion of particles in the vertical plane. They are slower than both P-waves and S-waves, but have high amplitudes and long durations, contributing to significant ground shaking. Rayleigh waves are usually in the frequency range from \qty{0.01}{\hertz} to \qty{1}{\hertz}.
    \item[Love waves] are surface waves that cause horizontal shear movements parallel to the Earth's surface and orthogonal to the direction of wave propagation. Love waves are usually in the frequency range from \qty{0.01}{\hertz} to \qty{1}{\hertz}. They are faster than Rayleigh waves but generally slower than P-waves and can be close in speed to S-waves. 
\end{description}
Seismic waves, especially the higher-frequency components of the seismic waves, attenuate as they travel through the Earth due to energy loss caused by scattering and absorption. Therefore, it is expected that distant earthquakes will be detectable only within the low-frequency range (\(\leq \qty{5}{\hertz}\)).

\subsubsection{Ocean-generated microseism}
    is continuous ground motion caused by the interaction of ocean waves with the seafloor. These microseisms are typically categorized into two main types:
\begin{description}
    \item[Primary microseism] is caused by the direct impact of ocean waves on the coast or the shallow seafloor. Primary microseisms are typically in the frequency range of about  \qty{0.05}{\hertz} to \qty{0.15}{\hertz} in the Atlantic.
        
    \item[Secondary microseism] is caused by the interaction of ocean waves with each other. The periods of these microseisms are about half as long as those of the primary microseisms, which corresponds to a frequency of about \qty{0.1}{\hertz} to \qty{0.3}{\hertz} in the Atlantic.
\end{description}

\subsubsection{Civilization noise} is ground movement caused by humans. The EuXFEL is located in Hamburg, a city with a population of almost \num{1.9} million inhabitants\cite{furschleswig}. Therefore, ground vibrations caused by human activity are also significant. By this we mean car and train traffic, industry, construction activities, but also cultural events\cite{diaz2020seismometers}. We expect these civilization-induced ground vibrations to be stronger during the day than at night due to the time of day and human activity. The reaction of people to major events such as concerts or sport events lead to ground movements. This has been observed at different occasions (e.g.,\,Taylor Swift concert\cite{bgspress2024quake}, American football game\cite{vidale2011one}). The EuXFEL is located near to Hamburg's Volksparkstadion and the Bahrenfelder Trabrennbahn, two locations where concerts and sport events often take place.

\section{Method}
\label{sec:method}

The optical synchronization system uses a feedback system for each individual link (see Figure~\ref{fig:lbsyncscheme}) which actively stabilizes the length of the link in a PLL using a PI controller. The phase difference \(\Phi_{\text{error}} \) between the pulse coming from the MLO and the pulse coming from the client is influenced by variations of the link length. The PI controller generates the adjustment for the piezo stretcher,  \(V_c(t)\), based on the determined phase difference:
\begin{equation}
    V_c(t) = K_p \cdot \Phi_{\text{error}}(t) + K_i \cdot \int_0^t \Phi_{\text{error}}(\tau) \, d\tau,
\end{equation}
where \(K_p\) corresponds to the proportional gain and \(K_i\) to the integral gain. Using a calibration factor \(c_{piezo}\), we calculate the compensated change in length of the link fiber. In addition to the piezo stretcher, the optical delay line is also controlled in regular system operation. We obtain the total distance changes to the path traveled by the optical pulse \(V_{out}(t) =  c_{piezo} \cdot V_c(t) + d_{odl}(t)\) by combining the paths of the optical delay line \(d_{odl}(t)\) and the delay introduced by the piezo stretcher. This value includes only the length changes compensated by the controller. The length changes which are not compensated by the controller can be observed in the phase difference \(\Phi_{\text{error}}\). We use the controller output \(V_{out}(t)\) to measure the seismic disturbances and the phase difference \(\Phi_{\text{error}}\) to determine the influence of disturbances on the optical synchronization system. 

\begin{figure*}[p]
  \centering
  \includegraphics[width=1\textwidth]{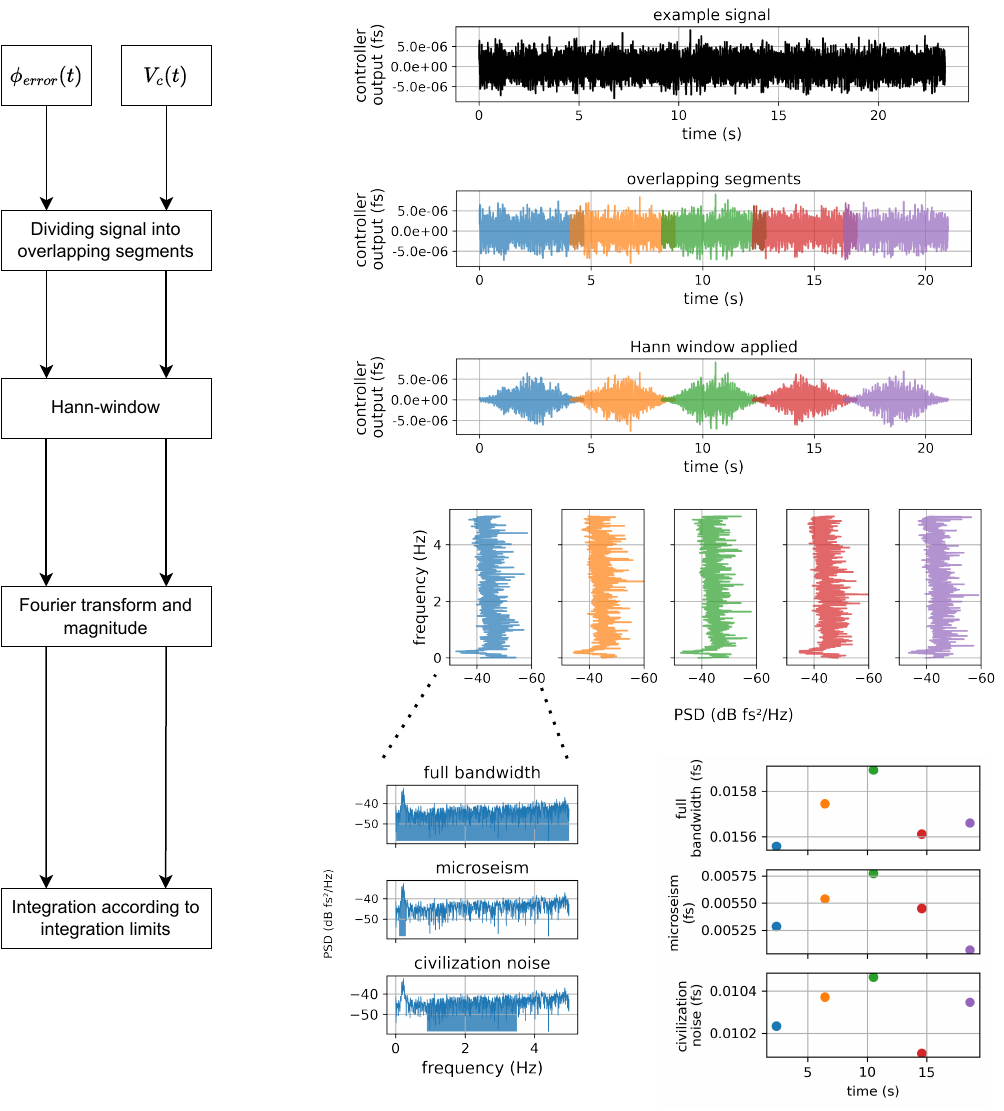}
  \caption{Overview of the method to analyze the impact of seismic activities on the optical synchronization system.}
  \label{fig:method}
\end{figure*}

The process to determine the impact of seismic activities from the respective time signals \(V_{out}\) and \(\Phi_{\text{error}}\) is shown in Figure~\ref{fig:method}. A discrete time signal \(x(n)\) of length \(N\) is divided into equally sized overlapping segments \(x_m\) of length \(L\), where \(m\) is the index of the respective segment. \textcolor{black}{ This segmentation is performed to focus the analysis on smaller, manageable time intervals, which allows for a detailed examination of the fluctuations caused by seismic disturbances and other noise sources. The overlap ensures that no information is lost at the boundaries of the segments, maintaining continuity in the analysis.} To avoid spectral leakage, the Hann window\cite{blackman1958measurement} is applied to each of the segments. The Hann-window function \(w(n)\) is defined as:
\begin{equation}
w(n) = 0.5 - 0.5\cos(\frac{2\pi n}{L-1}), \quad n=0, 1, \dots, L-1
\end{equation}
The windowed segment \(\ \tilde{x}_m(n)\) is the product of the original segment \(x_m(n)\) and the Hann-window \(w(n)\):
\begin{equation}
    \tilde{x}_m(n) = x_m(n) \cdot w(n), \quad n=0,1, \dots, L-1
\end{equation}
For each windowed segment, the Discrete Fourier Transform (DFT) is computed to transform the time-domain signal into the frequency domain. The frequency domain representation \(X_m\) for the \(m\)-th segment is given by:
\begin{equation}
    X_m(f) = \sum_{n=0}^{L-1} \tilde{x}_m(n) \cdot e^{-\frac{j \cdot 2\pi \cdot f \cdot n}{L}}, \quad f = 0, 1, \dots, L-1,
\end{equation}
where \(f\) denotes the frequency bin index and \(e^{-j \cdot 2\pi \cdot f \cdot n / L}\) is the complex exponential basis function. For efficient calculation, the DFT is calculated using the Fast Fourier Transform (FFT) algorithm\cite{cooley1965algorithm}. The Power Spectral Density (PSD) estimation \(P_{m}(f)\) of the \(m\)-th segment is obtained by calculating the absolute value of the Fourier-transform:
\begin{equation}
    P_{m}(f) = \lvert X_m(f) \rvert^2
\end{equation}
The PSD provides information on how the power of a time-series signal segment \(x_m\) is distributed over the frequency range. \textcolor{black}{This allows us to identify the frequency components associated with seismic disturbances, such as ocean-generated microseisms or civilization noise. By analyzing the PSD, we can track how the system's behavior fluctuates due to these disturbances.} The signals \(V_{out}\) and \(\Phi_{\text{error}}\) have units of femtoseconds. Therefore, the integration over a frequency interval results in an integrated jitter, which quantifies the fluctuations in the time signals caused by seismic activities.

The integrated jitter is calculated by integrating the PSDs over the frequency intervals resulting from the characteristics of the seismic activities, as described in Section~\ref{sec:seismics} and summarized in Table~\ref{tab:bandwiths}.

\begin{table}[H]
    \centering
    \caption{Frequency bandwidths of the seismic categories}
    \begin{tabular}{ccc}
    \toprule
    seismic category            & \(f_{start}\) (\si{\hertz}) & \(f_{end}\) (\si{\hertz})\\
    \midrule
    full bandwidth              &  \num{0.005} & \num{5}\\
    ocean-generated microseism  &  \num{0.1} & \num{0.3} \\
    civilization noise          &  \num{0.9} & \num{3.5} \\
    \bottomrule
    \end{tabular}
    \label{tab:bandwiths}
\end{table}

The integrated timing jitter of the \(m\)-th segment \(\sigma_{\Delta T, m}\) is calculated by integrating the PSD over a specified frequency interval:
\begin{equation}
    \sigma_{\Delta T, m} = \sqrt{\int_{f_{start}}^{f_{end}} P_{m}(f) df}
\end{equation}
The result of the method shown in Figure~\ref{fig:method} is that we get a jitter for each of the three bandwidths for the controller output and the controller input:
\begin{itemize}
    \item controller output - full bandwidth
    \item controller output - microseism
    \item controller output - civilization noise
    \item controller input - full bandwidth
    \item controller input - microseism
    \item controller input - civilization noise
\end{itemize}

The optical signal travels the distance twice, forward and backward. Therefore, the jitter values of the optical sync system are halved to ensure that the resulting jitter values correspond to the one-way tunnel length. The category-specific jitter describes the fluctuations of the signal caused by seismic activity. To determine the exact influence of seismic activity, the difference between the maximum and minimum timing jitter is calculated. The maximum jitter occurs when there is a strong disturbance due to seismic activity, while the minimum jitter represents the state of the system at rest, i.e.,\, when there is no or only negligible seismic activity. Furthermore, it should be noted that the earthquake effects and the effects of ocean-generated microseism overlap in the frequency domain. Therefore, to determine the jitter associated with ocean-generated microseism, we have cut out the time intervals of earthquake effects, as these would cause a significantly higher jitter than ocean waves. 

The data has been recorded during regular operation of the optical synchronization system. This means that the data set includes maintenance work, configuration changes, and online optimizations. These time intervals, which would also corrupt the evaluation of seismic activities, were also omitted. 

In addition to the optical synchronization system data, we use external data sources to validate and correlate the observed seismic disturbances. These include an earthquake database\cite{10.1785/0220160204}, local seismometer measurements from the EuXFEL injector building, the number of cars driving on a road above the EuXFEL tunnel\cite{traffic}, and sea level data measured by a buoy in the North Sea\cite{sealevel}. The external data help confirm the specific types of seismic events that influence the synchronization system and provide a broader context for understanding the observed fluctuations. For instance, civilization noise, which can be determined with the optical synchronization system, does not only depend on the cars on a single road. The traffic above the EuXFEL tunnel serves as a general indicator of human activity near the EuXFEL. Additionally, the turbulence measured by the North Sea buoy does not correspond exactly to ocean waves. However, we assume that increased ocean-generated microseism leads to an increased sea state and consequently greater ground motion at the EuXFEL, although the sea state is also influenced by other factors.

\section{Results}
\label{sec:results}

\begin{figure*}[b]
  \centering
  \includegraphics[width=0.9\textwidth]{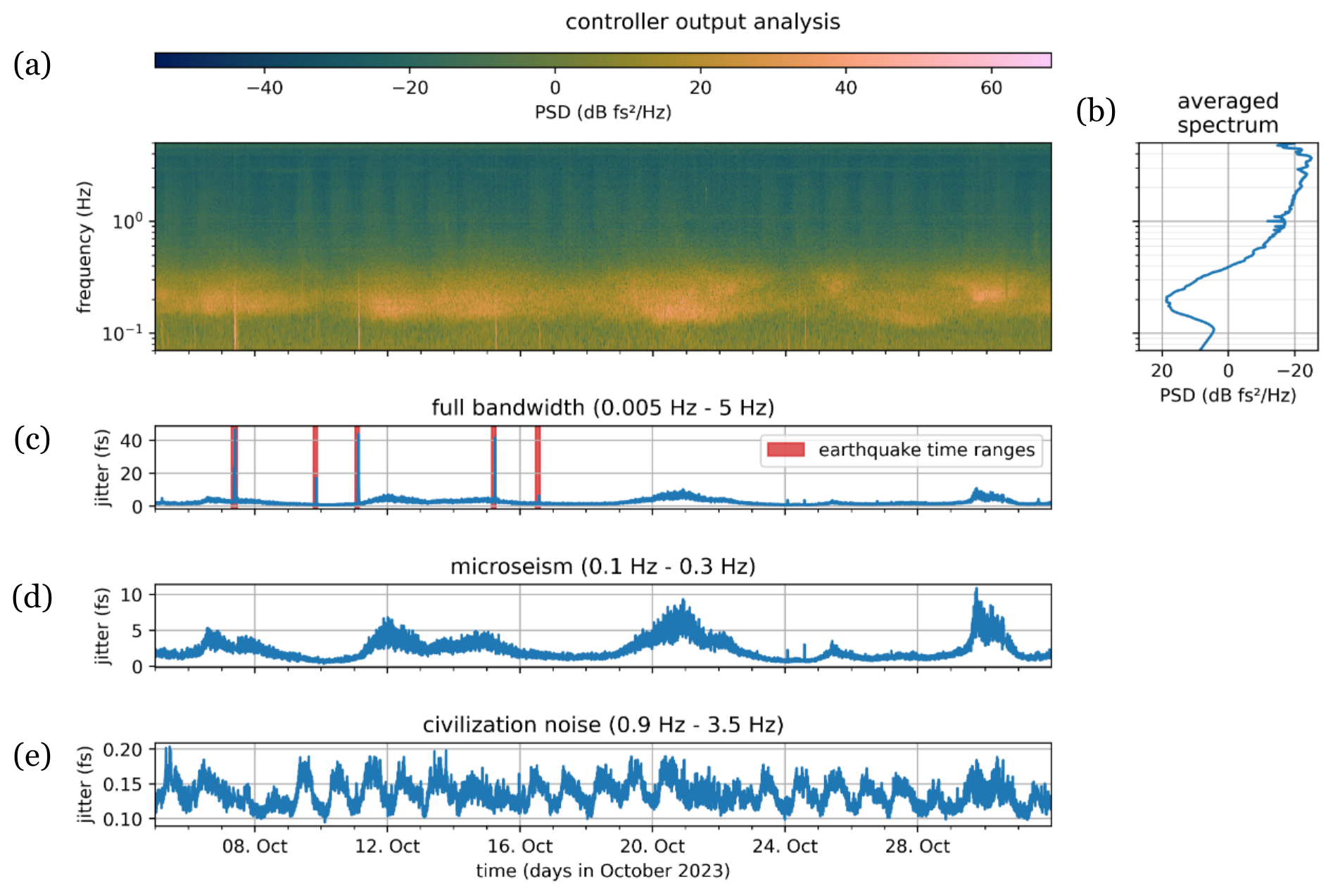}
  \caption{Results of analyzing \(V_{out}\) of the  \qty{3.5}{\km} link. (a) Spectrogram of October 2023. (b) Average PSD of October 2023. (c) Integrated jitter over the frequency range \qty{0.005}{\hertz} to  \qty{5}{\hertz}. (d) Integrated jitter over the frequency range \qty{0.1}{\hertz} to \qty{0.3}{\hertz}, with earthquake effects removed. (e) Integrated jitter over the frequency range \qty{0.9}{\hertz} to \qty{3.5}{\hertz}.}
  \label{fig:PSD-out-lsu-8.2}
\end{figure*}

For analyzing the seismic activities and the resulting disturbances in the optical synchronization system, control data from \num{12} links starting from the MLO are recorded at a data rate of \qty{10}{\hertz} during October \num{2023} using the associated data acquisition system\cite{grunhagen2023enhancing}. Windows of size \qty{200}{\second} were used to calculate the PSDs. Figure~\ref{fig:PSD-out-lsu-8.2} shows the analysis of the \(V_{out}(t)\) signal for the longest of the \num{12} recorded links, which leads to a client at a tunnel position of \qty{3.5}{\km} - representing the compensated length changes of the link. The spectrogram clearly shows all three categories of seismic activities. Specifically, noise triggered by ocean-generated microseism between \qty{0.1}{\hertz} and \qty{0.3}{\hertz}, and civilization noise between \qty{0.9}{\hertz} and \qty{3.5}{\hertz}, are visible. Seismic activities triggered by earthquakes were also detected during the data period.

Table~\ref{tab:earthquakes} summarizes the significant earthquake events that occurred in October \num{2023} and affected the optical synchronization system. The effects of these earthquakes are particularly noticeable in the low-frequency domain of the spectrogram. Earthquakes with magnitudes up to \num{6.7} led to integrated jitter spans of \qty{45}{\femto\second}, followed by ocean-generated microseism with jitter spans of up to \qty{10}{\femto\second}. In comparison, the jitter caused by civilization noise is significantly smaller, with spans of up to \qty{0.1}{\femto\second}.

\begin{table}[!ht]
    \centering
    \caption{October \num{2023} Earthquakes}
    \begin{tabular}{rcc}
    \toprule
    date              & Location & Magnitude \\
    \midrule
     7th October      &  Afghanistan & \num{6.3}  \\
     9th October      &  Poland      & \num{5}  \\
    11th October      &  Afghanistan & \num{6.3}  \\
    15th October      &  Afghanistan & \num{6.3}  \\
    16th October      &  Alaska      & \num{6.7}  \\
    \bottomrule
    \end{tabular}
    \label{tab:earthquakes}
\end{table}

\begin{figure}
  \centering
  \includegraphics[width=0.48\textwidth]{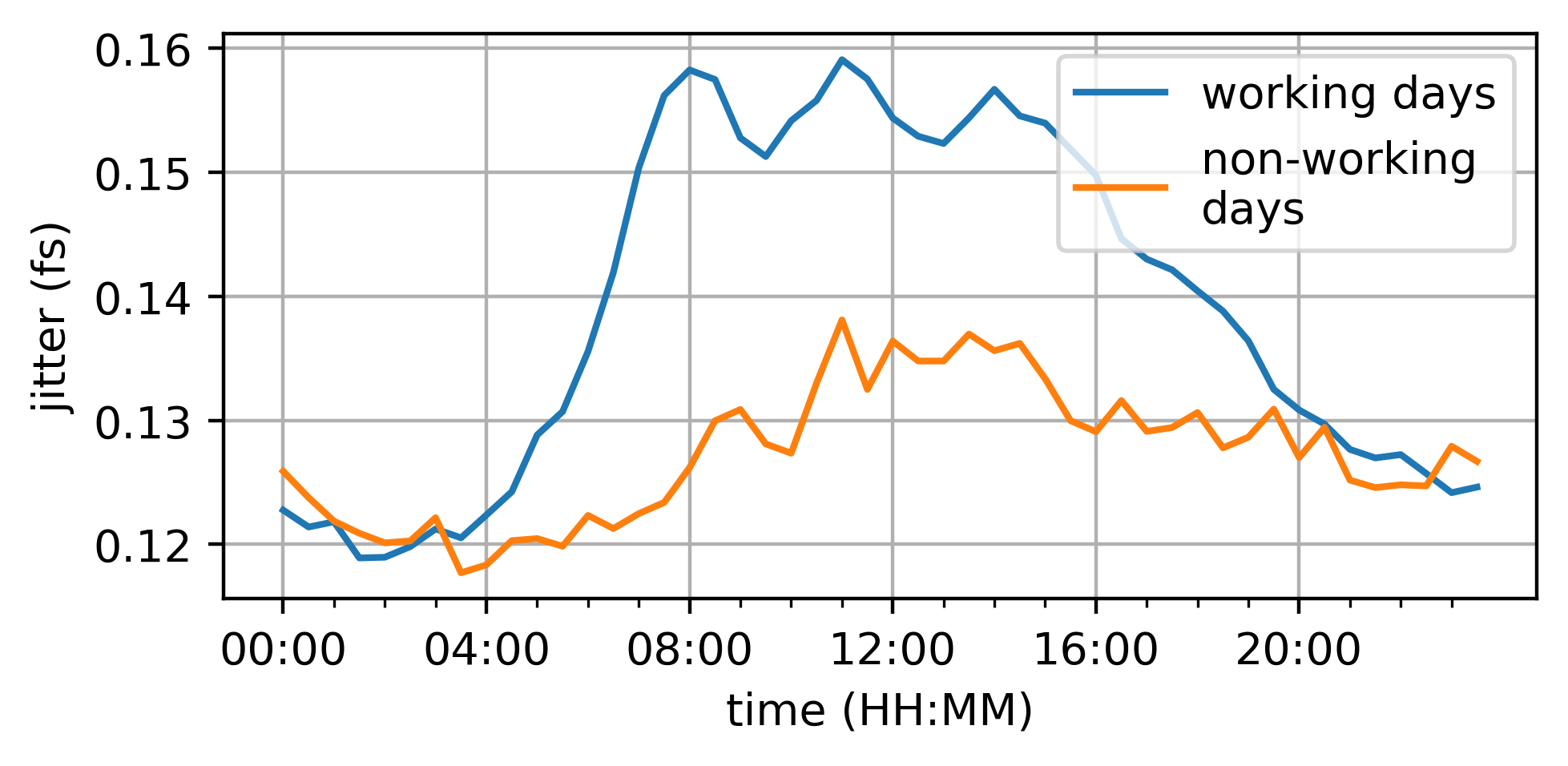}
  \caption{Comparison of civilization noise (0.9 Hz - 3.5 Hz) between working days and non-working days with respect to the time of day, averaged over \num{27} days.}
  \label{fig:daylijitter-lsu-8.2}
\end{figure}

The spectrogram and the integrated jitter of the civilization noise show a dependency between the ground movements and the time of day and weekday. Figure~\ref{fig:daylijitter-lsu-8.2} shows the integrated jitter averaged over a month (\qty{0.9}{\hertz} - \qty{3.5}{\hertz}) as a function of the time of day for the controller output. It is clearly visible that the time of day and whether it is a working day or not have an influence on the ground movements at the EuXFEL, which in turn affect the controller output of the optical synchronization system.

\begin{figure*}[!ht]
  \centering
  \includegraphics[width=1\textwidth]{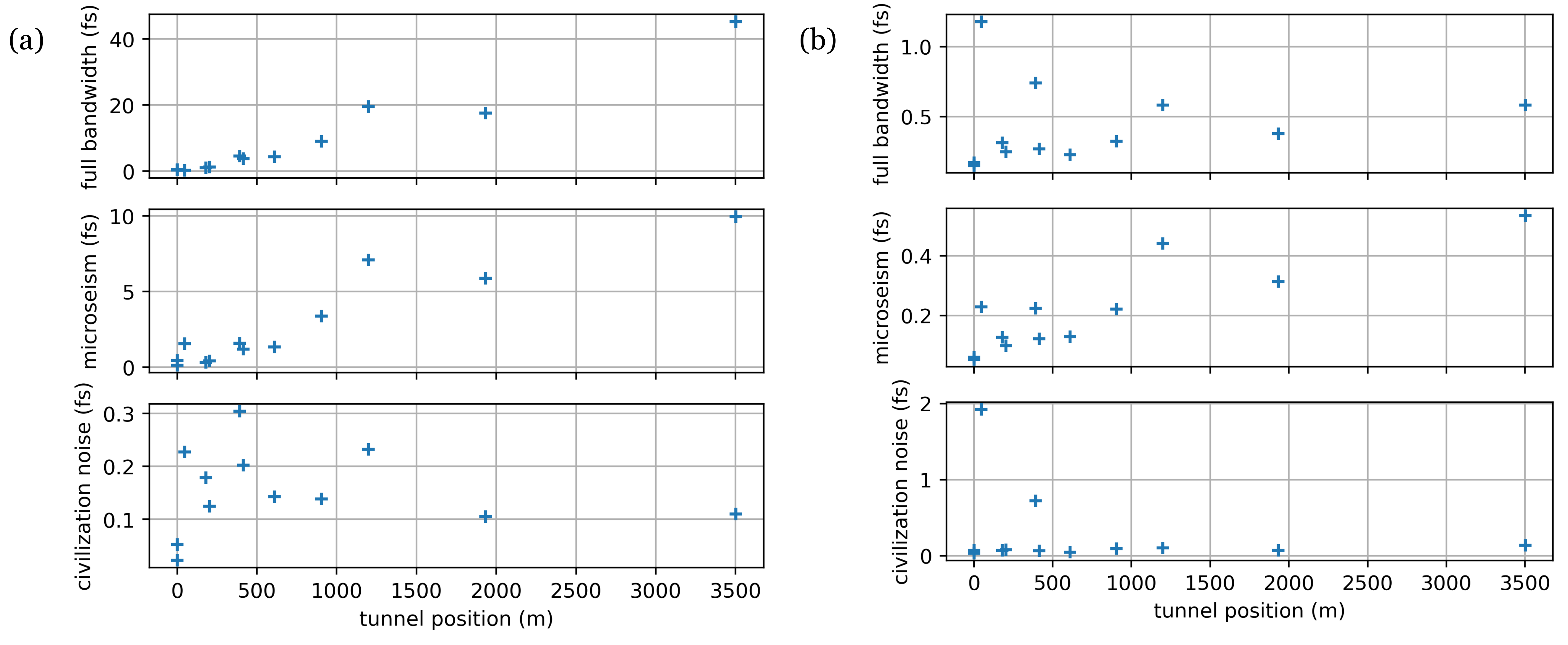}
  \caption{\textcolor{black}{Influence of seismic activities on optical links leading to different tunnel positions at EuXFEL. (a) The difference between the maximum and minimum integrated jitter values recorded over the month of October for each link, highlighting the variability in seismic impact. (b) The minimum integrated jitter observed during the same period, representing the baseline performance of the synchronization system under minimal seismic disturbance.}}
  \label{fig:jitters-length}
\end{figure*}

So far, only the results of the analysis on the \qty{3.5}{\km} link have been considered. The same analyses were also carried out for the remaining \num{11} links. Figure~\ref{fig:jitters-length} shows the tunnel position in relation to the spans of the three controller output integrated jitter categories. It can be seen that the spans of the earthquake jitter and of the ocean wave jitter increase with an increasing tunnel position.

Among the links, the \qty{1198}{\m} link stands out as an outlier, exhibiting unusually strong disturbances compared to neighboring links. This is likely due to the fact that the client associated with this link is situated directly beneath Rugenbarg, a heavily trafficked road. The increased disturbance can also be explained by the elevated civilization noise, which is strongest for tunnel positions between \qty{50}{\m} and \qty{1200}{\m}.

\begin{table*}[!b]
    \centering
    \caption{Spearman's correlation between the full bandwidth (full), microseism (micro), and civlization (civil) integrated jitters of the controller input (in), controller output (out), and external data sources}
    \begin{tabular}{ccccccc}
    \toprule
    data source                 & out - full                & out - micro         & out - civil         & in - full           & in - micro          & in - civil \\
    \midrule
    seismometer full bandwidth  & \num{0.6677}          & -             & -             & \num{0.5824}     & -              & - \\
    seismometer microseism      & -                     & \num{0.9173}  & -             & -                & \num{0.9264}   & - \\
    seismometer civilization    & -                     & -             & \num{0.6480}  & -                & -              & \num{-0.0600} \\
    sea level                   & -                     & \num{0.6162}  & -             & -                & \num{0.5996}   & - \\
    car count                   & -                     & -             & \num{0.6916}  & -                & -              & \num{-0.1230} \\
    \bottomrule
    \end{tabular}
    \label{tab:external}
\end{table*}

The variation in the impact of seismic activities across the links can be attributed to several factors, such as optical link length, required optical power at the client, and the specific environmental conditions at different tunnel positions. Due to the these factors the respective noise floors of the integrated jitters deviate.

The calculated spans of the civilization noise jitters are close to the noise floor levels for some links. Therefore, it is not clear from these results that the links leading to clients with tunnel positions between \qty{50}{\m} and \qty{1200}{\m} are most affected by civilization noise. The full-bandwidth jitter and the ocean-generated microseism jitter are significantly larger than the background noise, indicating that the influence of earthquakes and ocean-generated microseism increases with increasing tunnel position.

\begin{figure*}[!ht]
  \centering
  \includegraphics[width=1\textwidth]{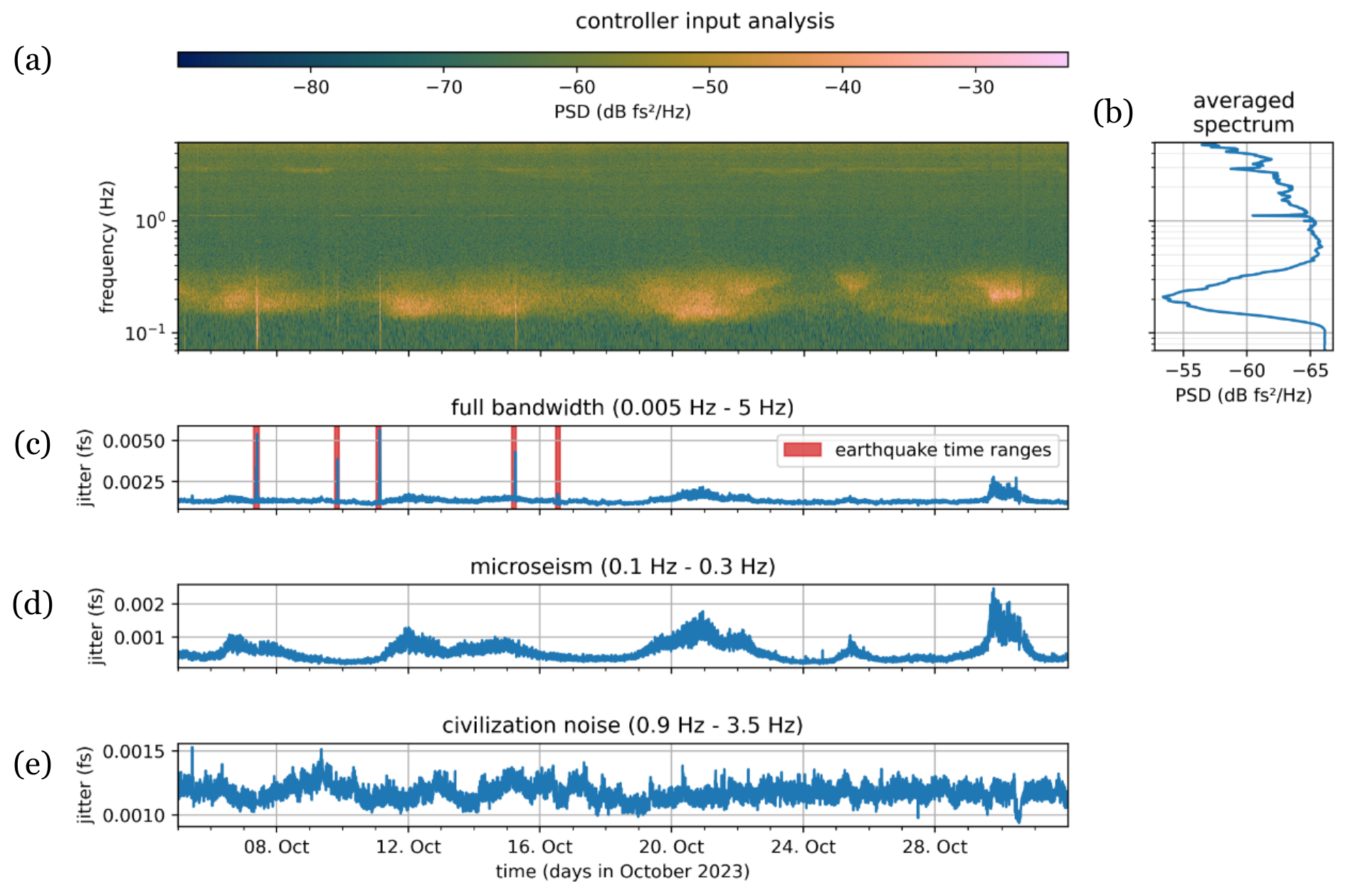}
  \caption{Results of analyzing \(\Phi{error}\) of the  \qty{3.5}{\km} link. (a) Spectrogram of October 2023. (b) Average PSD of October 2023. (c) Integrated jitter over the frequency range \qty{0.005}{\hertz} to  \qty{5}{\hertz}. (d) Integrated jitter over the frequency range \qty{0.1}{\hertz} to \qty{0.3}{\hertz}, with earthquake effects removed. (e) Integrated jitter over the frequency range \qty{0.9}{\hertz} to \qty{3.5}{\hertz}.}
  \label{fig:PSD-in-lsu-8.2}
\end{figure*}

The analysis of the controller input \(\Phi_{\text{error}}\) was conducted similarly to the controller output analysis. Figure~\ref{fig:PSD-in-lsu-8.2} shows the results of analyzing the controller input of the link leading to the a client with a tunnel position of \qty{3.5}{\km}. Earthquakes lead to an integrated jitter span of more than \qty{4}{\atto\second}, while ocean-generated microseism causes jitter spans up to \qty{2}{\atto\second}. Civilization noise, however, cannot be distinguished from the background noise. The averaged controller input PSD shows that the largest disturbance occurs at approximately \qty{0.2}{\hertz}, which corresponds to the frequency range of ocean-generated microseism.

\subsection{Comparison to external data}
In addition to the optical synchronization system control data, we use external data to validate and correlate the observed seismic disturbances. These include an earthquake database, data from a seismometer located in the EuXFEL injector building, traffic data from a road above the EuXFEL tunnel, and sea level measurements from a buoy in the North Sea. These external sources help to provide a broader context for understanding the observed fluctuations and confirm the specific types of seismic events that influence the synchronization system. Table~\ref{tab:external} summarizes the correlation between the system integrated jitter of the controller I/O and external data sources with the Spearman's rank coefficient\cite{zwillinger1999crc}.

\begin{figure*}[!ht]
  \centering
  \includegraphics[width=0.93\textwidth]{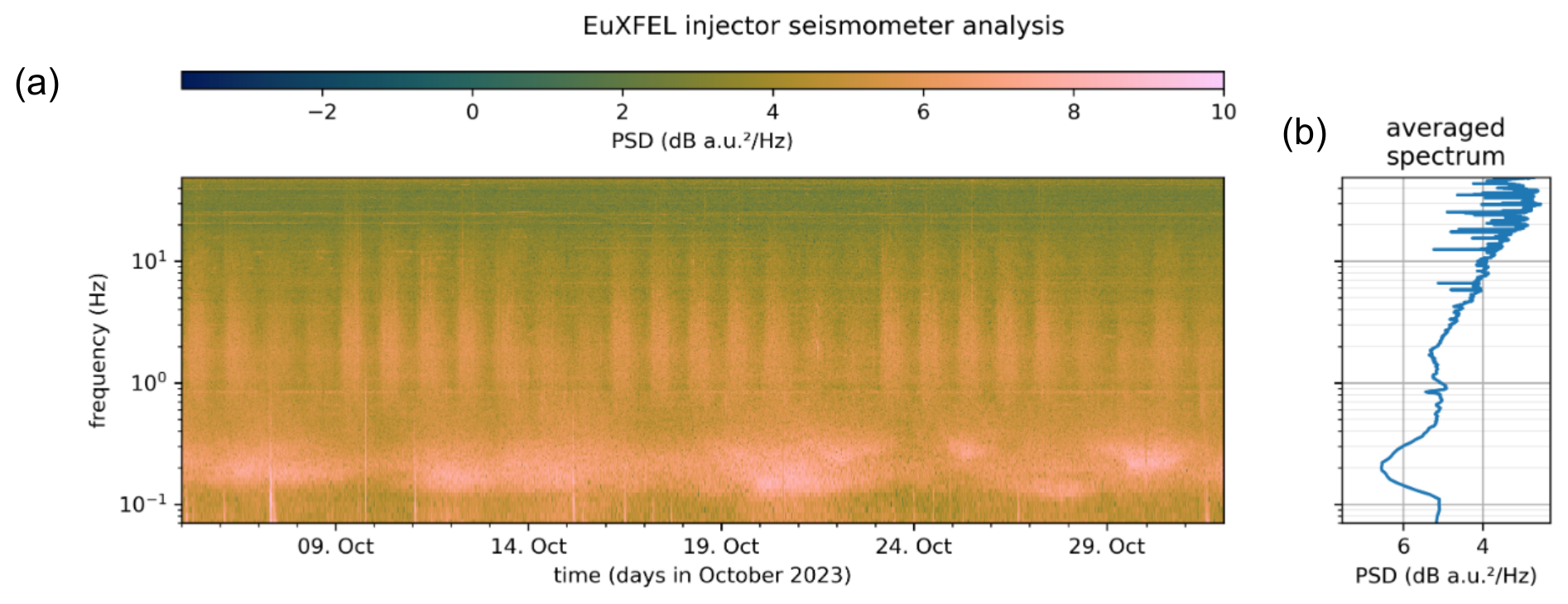}
  \caption{\textcolor{black}{Spectrogram (a) and averaged PSD (b) of seismometer data in EuXFEL tunnel direction.}}
  \label{fig:seismometer}
\end{figure*}

\begin{figure*}[!ht]
  \centering
  \includegraphics[width=\textwidth]{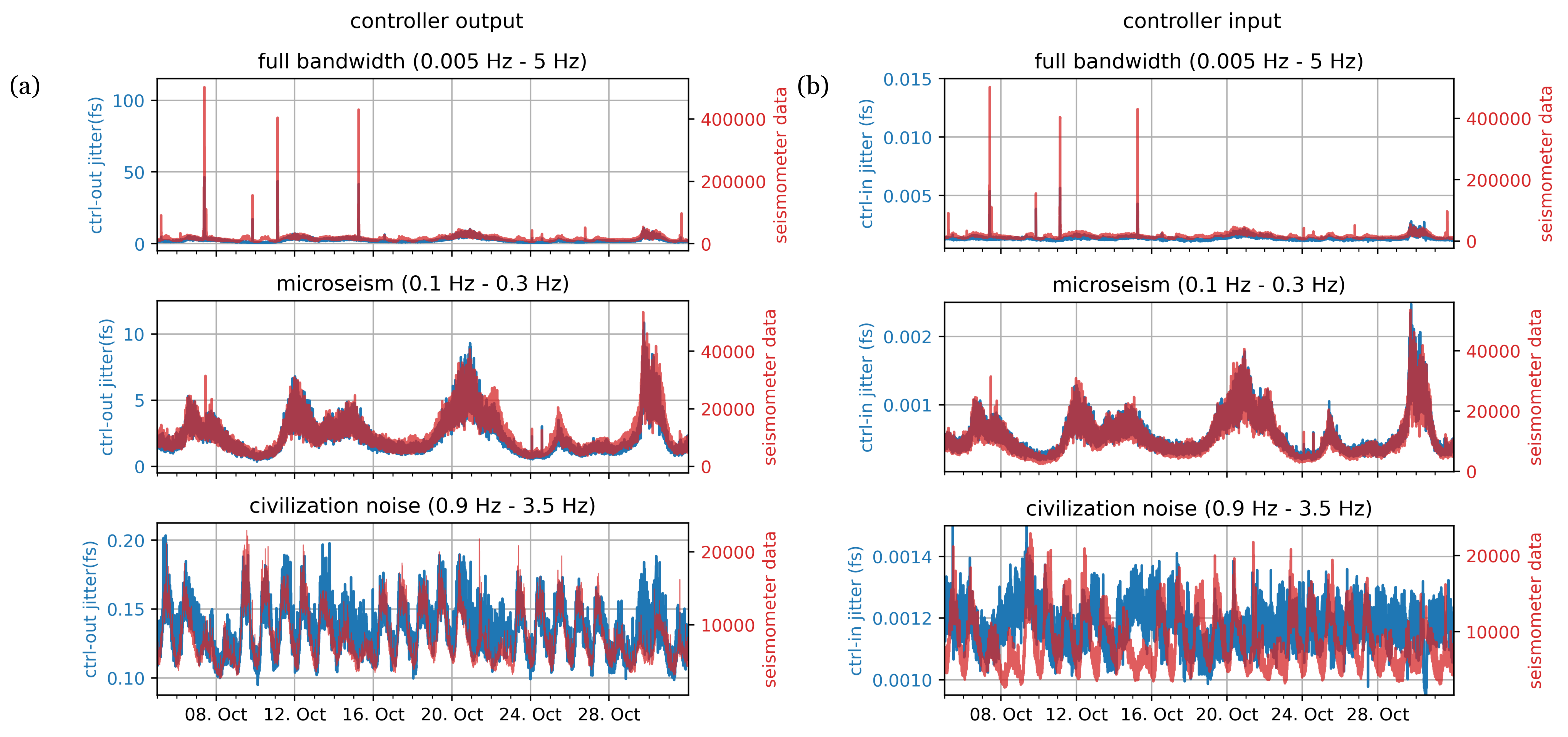}
  \caption{Comparison between the integrated jitters of controller output (a), input (b), and seismometer data.}
  \label{fig:compare-ctrlout-seismometer}
\end{figure*}

The data measured by the seismometers at the EuXFEL injector is shown in Figure~\ref{fig:seismometer} and Figure~\ref{fig:compare-ctrlout-seismometer} shows the integrated jitters of the controller I/O and the corresponding integrated jitters of the seismometer data.

When considering the full frequency domain, no linear relationship is observed between the seismometer data and the controller data. During earthquakes, the seismometers values are significantly larger than the controller output values, unlike the microseism and civilization noise, which show better alignment. For the microseism frequency range, the integrated jitter of both the controller output and input strongly correlates with the seismometer data, as indicated by Spearman coefficients of \num{0.9173} and \num{0.9264}, respectively. In contrast, the full bandwidth integrated jitter shows moderate correlations of \num{0.6677} (controller output) and \num{0.5824} (controller input).

For civilization noise, the time-of-day dependence is more pronounced in the seismometer data than in the controller output, as reflected in a Spearman coefficient of \num{0.648}. However, no significant time-of-day or day-of-week dependence is observed in the controller input, resulting in a low correlation coefficient (\num{-0.0600}). This suggests that the controller effectively attenuates low-amplitude vibrations caused by civilization noise.

\begin{figure}[!ht]
  \centering
  \includegraphics[width=0.48\textwidth]{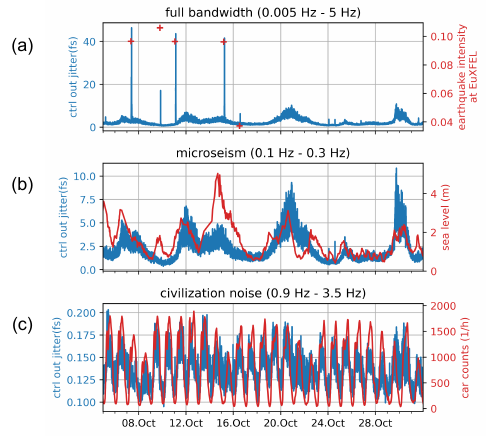}
  \caption{Comparison between the integrated jitters of controller output, earthquake intensities (a), sea level (b), and car counts (c).}
  \label{fig:out-external}
\end{figure}

Figure~\ref{fig:out-external} illustrates the relationship between the controller output and external data for each jitter category.
\begin{itemize}
    \item \textcolor{black}{\textbf{Earthquakes:} Figure~\ref{fig:out-external}a compares the full bandwidth jitter to estimated earthquake intensities \(I\) at EuXFEL, calculated using \(I \approx \nicefrac{10^M}{d^2}\), where \(M\) is the magnitude and \(d\) the distance from the EuXFEL to the earthquake epicenter. Despite simplifying assumptions, such as neglecting soil conditions and wave types, the correlation between jitter peaks and earthquake events is clear. However, the specific jitter impact cannot be deduced solely from magnitude and distance.
    }
    \item \textbf{Ocean-Generated Microseism:} Figure~\ref{fig:out-external}b compares the microseism jitter to sea level measurements from a North Sea buoy. Periods of strong correlation (e.g., 6–9 Oct, 11–13 Oct) are evident, but discrepancies (e.g., 5–6 Oct) highlight the influence of additional factors. Spearman coefficients of \num{0.6162} (controller output) and \num{0.5996} (controller input) indicate a moderate correlation.
    \item \textbf{Civilization Noise:} Figure~\ref{fig:out-external}c compares the civilization jitter with the traffic data. A strong correlation is observed between the controller output and the number of vehicles (\num{0.6916}), while the controller input shows no significant correlation (\num{-0.1230}). This indicates that the system is capable of mitigating low-level vibrations caused by traffic.
\textcolor{black}{
It is important to note that the seismometer used in this study is installed in the injector building, which allows it to detect higher-frequency civilization noise up to \qty{20}{\hertz}. In contrast, the data from the optical synchronization system reflects only ground movements at the tunnel level, which is loosely connected to above-ground structures like the injector building. Consequently, above-ground noise is significantly more attenuated in the measurement data from the optical synchronization system compared to the seismometer data.
}
\end{itemize}

In summary, the external data comparison validates the optical synchronization system’s sensitivity to seismic events and provides insights into the nature of these disturbances. While the system effectively mitigates low-level vibrations, further research is needed to better understand relationships, particularly for earthquake-induced disturbances. These findings highlight the importance of integrating external data sources for comprehensive system analysis and improvement.

\subsection{Ground Movements by Concerts}
\begin{figure*}
  \centering
  \includegraphics[width=1\textwidth]{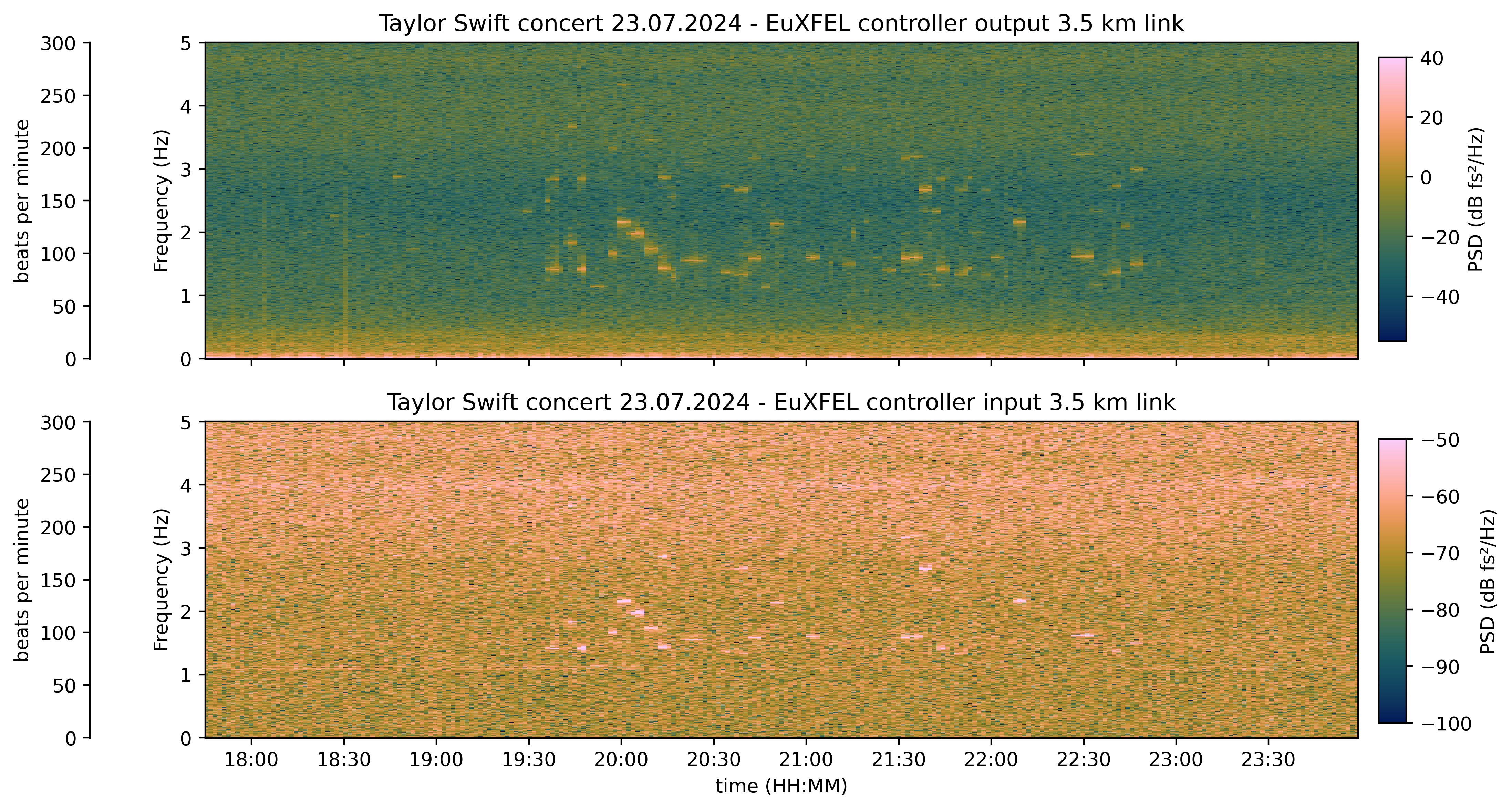}
  \caption{Spectrogram of Controller I/O during Taylor Swift concert.}
  \label{fig:taylor-outin}
\end{figure*}

The EuXFEL Injector is located in about \qty{2}{\km} distance from the Volksparkstadion in Hamburg. Figure~\ref{fig:taylor-outin} shows the spectrograms of the controller input and the controller output during a Taylor Swift concert at the Volksparkstadion on July 23, 2024. We can detect excitations in both the controller output and the controller input. These excitations correspond to the Beats Per Minute (BPM) of the songs played. For example, the song ‘lovestory’ was played with BPMs of \num{119} at 20:00, ‘shake it off’ with BPMs of \num{160} at 21:45. At these times, excitation frequencies of \qty{1.98}{\hertz} and \qty{2.67}{\hertz} were observed in the controller signals. This analysis demonstrates that the optical synchronization system is sensitive enough to detect the frequencies or BPMs of the songs played at the concert.

\section{Conclusion}
\label{sec:conclusion}
The optical synchronization system is a highly accurate and sensitive technology capable of detecting even the smallest seismic disturbances. This study analyzes the controller input and output signals of phase-locked loops (PLLs) from 12 length-stabilized optical links spanning up to \qty{3.5}{\km} within the EuXFEL tunnel. Our analysis revealed the extent of seismic activity at the facility and its impact on the optical synchronization system, with a particular focus on compensated and uncompensated disturbances in the low-frequency range.

\textcolor{black}{
The methodology presented in this study offers a highly effective approach to analyzing seismic impacts on optical synchronization systems. By combining controller I/O data with real-time seismic measurements, we capture seismic effects on system performance. This method not only provides accurate, time-resolved insights but also enables a detailed understanding of how various types of seismic activity influence synchronization. This level of analysis has not been widely explored in previous studies, making our approach a valuable contribution to the field.
}

Our analysis reveals that seismic disturbances from earthquakes, ocean-generated microseism, and civilization noise, affect system performance, with earthquakes causing the greatest disruption, followed by ocean-generated microseism and civilization noise. While ocean-generated microseism and civilization noise are persistent, earthquakes are infrequent but cause significant fluctuations in the system. Our results show that seismic effects triggered by earthquakes and ocean-generated microseism, intensify with increasing tunnel position. Ground movements caused by civilization noise are especially significant in areas where the EuXFEL tunnel passes beneath residential neighborhoods with heavy street traffic.

A key takeaway from this study is the strong correlation between seismic activity and fluctuations in the optical synchronization system. \textcolor{black}{However, further work is needed to establish direct relationships between seismic disturbances and observables such as x-ray photon or electron arrival times. Future research should also explore additional attributes of seismic events, such as the propagation direction of earthquake waves relative to the EuXFEL tunnel, as these factors may play a critical role.} Furthermore, optimizing the control loop to mitigate low-frequency disturbances remains a priority, as these effects have the most significant impact during normal operations.

In summary, the optical synchronization system exhibits remarkable precision, capable of detecting and attenuating \qty{99.99}{\percent} of disturbances caused by seismic activity. This capability not only ensures stable operation of the EuXFEL but also provides valuable insights into seismic phenomena.

\appendix{}

\sloppy{}

\begin{center}\textbf{Acknowledgment}\end{center}
We acknowledge the support by DASHH (Data Science in Hamburg - Helmholtz Graduate School for the Structure of Matter) with the Grant-No. HIDSS-0002.

\bibliographystyle{unsrt}
\bibliography{hpl-sample}




\end{document}